\documentclass[twocolumn,amsmath,amssymb,aps,prl]{revtex4-1}
\usepackage[dvips]{graphicx}
\usepackage{dcolumn}
\usepackage{bm}
\usepackage{float}
\usepackage{hyperref}
\usepackage{color}
\usepackage{cancel}
\usepackage{bmpsize}
\usepackage{amsmath}
\usepackage{amssymb}
\usepackage{natbib}

\newcommand{\ket}[1]{\left| #1\right\rangle}
\newcommand{\braket}[2]{\left\langle
#1\vphantom{#2}\right|\left.#2\vphantom{#1}\right\rangle}

\newcommand{\be}[0]{\begin{equation}}
\newcommand{\ee}[0]{\end{equation}}

\newcommand{\lra}\simeq
\newcommand{\eeqref}[1]{Eq.~(\ref{#1})}

\definecolor{linkcolor}{rgb}{0.9,0,0}
\definecolor{citecolor}{rgb}{0,0.6,0}
\definecolor{urlcolor}{rgb}{0,0,1}
\hypersetup{
	colorlinks, linkcolor={linkcolor},
	citecolor={citecolor}, urlcolor={urlcolor}
}

\begin{document}

\title{Breeding the optical Schr\"odinger's cat state}

\author{Demid V. Sychev$^{*1}$, Alexander E. Ulanov$^{*1,2}$, Anastasia A. Pushkina$^{1,3}$, Matthew W. Richards$^{4}$, Ilya A. Fedorov$^{*1,3}$, and A. I. Lvovsky$^{1,3,4}$}

\affiliation{$^1$Russian Quantum Center, 100 Novaya St., Skolkovo, Moscow 143025, Russia}
\affiliation{$^2$Moscow Institute of Physics and Technology, 141700 Dolgoprudny, Russia}
\affiliation{$^3$P. N. Lebedev Physics Institute, Leninskiy prospect 53, Moscow 119991, Russia}
\affiliation{$^4$Institute for Quantum Science and Technology, University of Calgary, Calgary AB T2N 1N4, Canada}
\affiliation{$^*$These authors contributed equally to the work.} 

\email{LVOV@ucalgary.ca}
\date{\today}

\begin{abstract}

Superpositions of macroscopically distinct quantum states, introduced in Schr\"{o}dinger's famous Gedankenexperiment, are an epitome of quantum ``strangeness" and a natural tool for determining the validity limits of quantum physics. The optical incarnation of Schr\"{o}dinger's cat --- the superposition of two opposite-amplitude coherent states --- is also the backbone of quantum information processing in the continuous-variable domain. However, existing preparation methods limit the amplitudes of the component coherent states by about 2, which curtails the state's usefulness for fundamental and practical applications.
Here we produce higher-amplitude optical Schr\"odinger's cats from two such states of lower amplitudes. The protocol consists in bringing the initial states into interference on a beamsplitter and a subsequent heralding quadrature measurement in one of the output channels. In the experiment, we convert a pair of negative squeezed Schr\"odinger's cat states of amplitude 1.25 to a single positive Schr\"odinger's cat of amplitude 2.15 with success probability of $\sim 0.2$. This amplitude is comparable to the highest values obtained for this state in any physical system. Our technique can be realized in an iterative manner, in principle allowing creation of Schr\"{o}dinger's cat states of arbitrarily high amplitude.

\end{abstract}

\maketitle

\vspace{10 mm}

In Schr\"{o}dinger's proposal, life and death of a cat are entangled with the state of a decaying atom, resulting in a macroscopic quantum superposition state \cite{Schroedinger1935}. This setting, originally used as a metaphor to demonstrate the absurdity of the newborn quantum theory in the macroscopic domain, remained a matter of thought experiments for half a century.
As quantum physics matured, this paradox was revisited; nowadays, Schr\"odinger's cat (SC) is being emulated in diverse physical systems. It is expected to help answering a fundamental question \cite{Haroche2013, Wineland2013, Arndt2014}: at what degree of macroscopicity, if any, does the world stop being quantum?

In optics, the SC state corresponds to a superposition of coherent states $\ket{\pm \alpha}$, which are considered the most classical of all states of light \cite{Leonhardt}. In such a superposition, the fields of the electromagnetic wave point in two opposite directions at the same time, resembling the superposition of dead and alive states of a cat in the original concept. The size parameter in this case is the amplitude $\alpha$. In order for the SC to be macroscopic, $\alpha$ has to be much larger than the quantum uncertainty $1/\sqrt{2}$ of the position observable in the coherent state \cite{Leonhardt, Leggett2002, Lvovsky2013}.

Apart from the fundamental interest, optical SC states are shown to be useful in applied quantum science. They can serve as a basis for quantum computation \cite{Ralph2003, FaultTolerant}, metrology \cite{QuantumMetrology}, teleportation and cryptography \cite{Near-deterministic, QuantumRepeater, LongDistance}. Most of these applications require the involved coherent states to be of reasonably high amplitudes. For example, encoding a qubit in the coherent state basis $\ket{\pm \alpha}$ is practical only when these states are nearly orthogonal,i.e.~when $\alpha \gtrsim 2$  \cite{Ralph2003}. Although a fault-tolerant quantum computation scheme optimised at $\alpha\sim1.6$ has been proposed, it requires a significant resource overhead \cite{FaultTolerant}.
Optical SC states which are currently available in experiment are yet to meet this condition. To the best of our knowledge, the record-size optical SCs have amplitudes $\alpha=1.73-1.84$, with fidelities on a scale of $67-88\%$ \cite{Minimal, ReverseHong}.

The above motivation to build optical SC states inspired significant experimental strive \cite{Minimal, ReverseHong, Ourjoumtsev2006, Nielsen2006, Wakui2007, Ourjoumtsev2009, Gerrits2010, Takahashi2013, Dong2014, PhotonNumber, Iterative}. Most of the existing experiments are based on photon subtraction from the squeezed vacuum state \cite{Ourjoumtsev2006, Nielsen2006, Wakui2007, Ourjoumtsev2009, Gerrits2010, Takahashi2013, Dong2014} or on quantum-state engineering  within the subspace of three lower Fock states \cite{Minimal, ReverseHong}. The state obtained by these methods approximates SCs reasonably well only for relatively small amplitudes. Better performance is offered by the method of preparing SCs from multiphoton Fock states \cite{PhotonNumber}, but these states are themselves difficult to prepare in a  reliable and scalable fashion.

In this work, we implement an alternative for the direct-preparation approach mentioned above. Our technique probabilistically converts a pair of SCs into a single SC state whose amplitude is greater than that of the initial ones  by a factor of $\sqrt{2}$. Our method can be applied in an iterative manner \cite{Amine1, Iterative}, allowing one to prepare an SC state of, in principle, any desirable amplitude --- given that sufficiently many initial SCs are available at the inception stage.

\textbf{Concept.}
The idea of the method has been proposed by Lund {\it et al.} \cite{LJRK} and further developed by Langhaout {\it et al.} \cite{Amine1}. Let the initial SC state be a superposition of coherent states of real amplitude $\alpha$:
\begin{subequations}
\label{eq1}
\begin{eqnarray}
\label{eq1a}
\ket{{\rm SC}_+\left[\alpha\right]} =\mathcal N (\ket{\alpha}+\ket{-\alpha}) \\
\label{eq1b}
\ket{{\rm SC}_-\left[\alpha\right]} = \mathcal N(\ket{\alpha}-\ket{-\alpha}),
\end{eqnarray}
\end{subequations}
where $\mathcal N$ is the normalization factor. Suppose a pair of identical, either positive or negative, states (\ref{eq1}) is put to interference on a symmetric beamsplitter. Let the relative phase of the inputs be such that equal coherent states  interfere constructively in its output mode 1 and destructively in mode 2, whereas opposite-amplitude coherent states behave in the converse fashion. The two-mode output state of the beam splitter is then \cite{Amine1}
\begin{equation}
\label{eq2}
\ket{\Psi}_{12}=\mathcal N\left(\ket{{\rm SC}_+[\sqrt{2}\alpha]}_1\ket{0}_2 \pm \ket{0}_1\ket{{\rm SC}_+[\sqrt{2}\alpha]}_2\right),
\end{equation}
where $\ket{0}$ is the vacuum state. If we now perform a measurement on mode 2 to distinguish the states $\ket 0$ and $\ket{{\rm SC}_+[\sqrt{2}\alpha]}$, and detect the vacuum state, mode 1 will collapse onto the positive SC state of  amplitude $\sqrt{2}\alpha$ \cite{LJRK}.

The required conditioning can be realized by homodyne measurement of the position quadrature in mode 2 and looking for the null result. Indeed, the wavefunction of state $\ket{{\rm SC}_+[\sqrt{2}\alpha]}_2$ is a sum of two Gaussians of width $1/\sqrt 2$ centered respectively at $X=\pm\alpha\sqrt 2$, whereas the wavefunction of the vacuum state is the same Gaussian centered at $X=0$. Therefore, for $\alpha\gg 1$, the probability of observing $X=0$ is much higher in the vacuum state than in state  $\ket{{\rm SC}_+[\sqrt{2}\alpha]}$. A precise calculation yields 
\begin{align}
\label{eq3}
\ket{\psi}_{1} & = \, _2\braket{X=0}{\! \Psi}_{12} \\
&=\mathcal N\left( \ket{{\rm SC}_+[\sqrt{2}\alpha]}_1 \pm \dfrac{\ket{0}_1}{\sqrt{(1+e^{4\alpha^2})/2}}\right).
\end{align}
For the initial SC amplitude $\alpha=1$, which is easily attained by present technology \cite{Minimal, ReverseHong, Ourjoumtsev2006, Gerrits2010, Takahashi2013, Dong2014, Iterative}, state (\ref{eq3}) has fidelity of 99\% with the ideal $\ket{{\rm SC}_+[\sqrt{2}]}_1$ state. As the weight of the vacuum component in (\ref{eq3}) decreases with $\alpha$, subsequent stages of amplification are possible with even higher fidelities (throughout this paper, the fidelity between the states with density operators $\rho_1$ and $\rho_2$ is defined as $F={\rm Tr}\left[(\sqrt\rho_1\rho_2\sqrt\rho_1)^{1/2}\right]$). 

\textbf{Experiment and results.}
\begin{figure}[t]
	\includegraphics[width=\columnwidth]{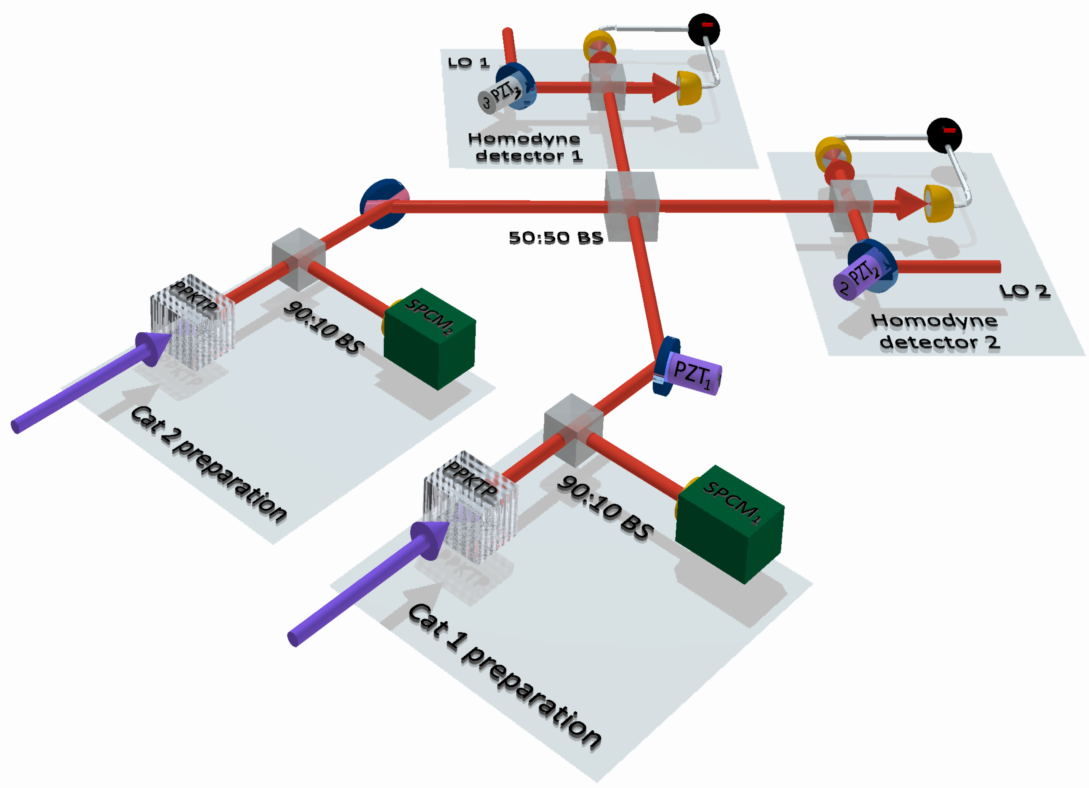}
	\caption{Scheme of the experiment. A squeezed vacuum state is generated in each input channel via degenerate parametric down-coversion (PPKTP). A small portion of the photon flux from each squeezer is directed to a single photon counting module (SPCM); preparation of the two initial negative SC states is heralded by simultaneous clicks of SPCMs 1 and 2. The resulting SC states interfere on a 50:50 beamsplitter (BS). Optical homodyne tomography of the state in output mode 1 of the BS is performed conditioned on the near-zero result of the position quadrature measurement in output mode 2. LO: local oscillator. PZT: piezoelectric transducer.
	}
	\label{f1}
\end{figure}
The experimental setup for realizing the above protocol is shown in Fig.~\ref{f1}.
The initial SCs in two spatially distinct light modes are generated by photon subtraction from squeezed vacuum states \cite{Dakna,Ourjoumtsev2006}. Since the squeezed vacuum is approximated by a positive SC (\ref{eq1a}), the photon annihilation flips the sign in front of the $\ket{-\alpha}$ component, converting the state to a negative, squeezed SC (\ref{eq1b}) of a larger amplitude \cite{Ourjoumtsev2006, Nielsen2006}. The obtained states are characterized by optical homodyne tomography \cite{Lvo2009, Lvo2004}.
Corrected for a total quantum efficiency of 50\% (see Methods), they have fidelity of 93\% with an ideal state $\ket{{\rm SC}_- \left[1.25\right]}$, squeezed by 1.73 dB. The Wigner functions of the experimentally reconstructed states and the best fit state are shown in Fig.~\ref{f2}(ii). Note that, throughout this paper, we follow the ideology of Ref.~\cite{PhotonNumber} by fitting our experimentally acquired states by \emph{squeezed} SCs. The rationale behind this convention is that squeezing  does not affect the macroscopicity of the superposition \cite{LeeJeong} and can be undone by a local unitary operation.

\begin{figure}[h]
	\includegraphics[width=\columnwidth]{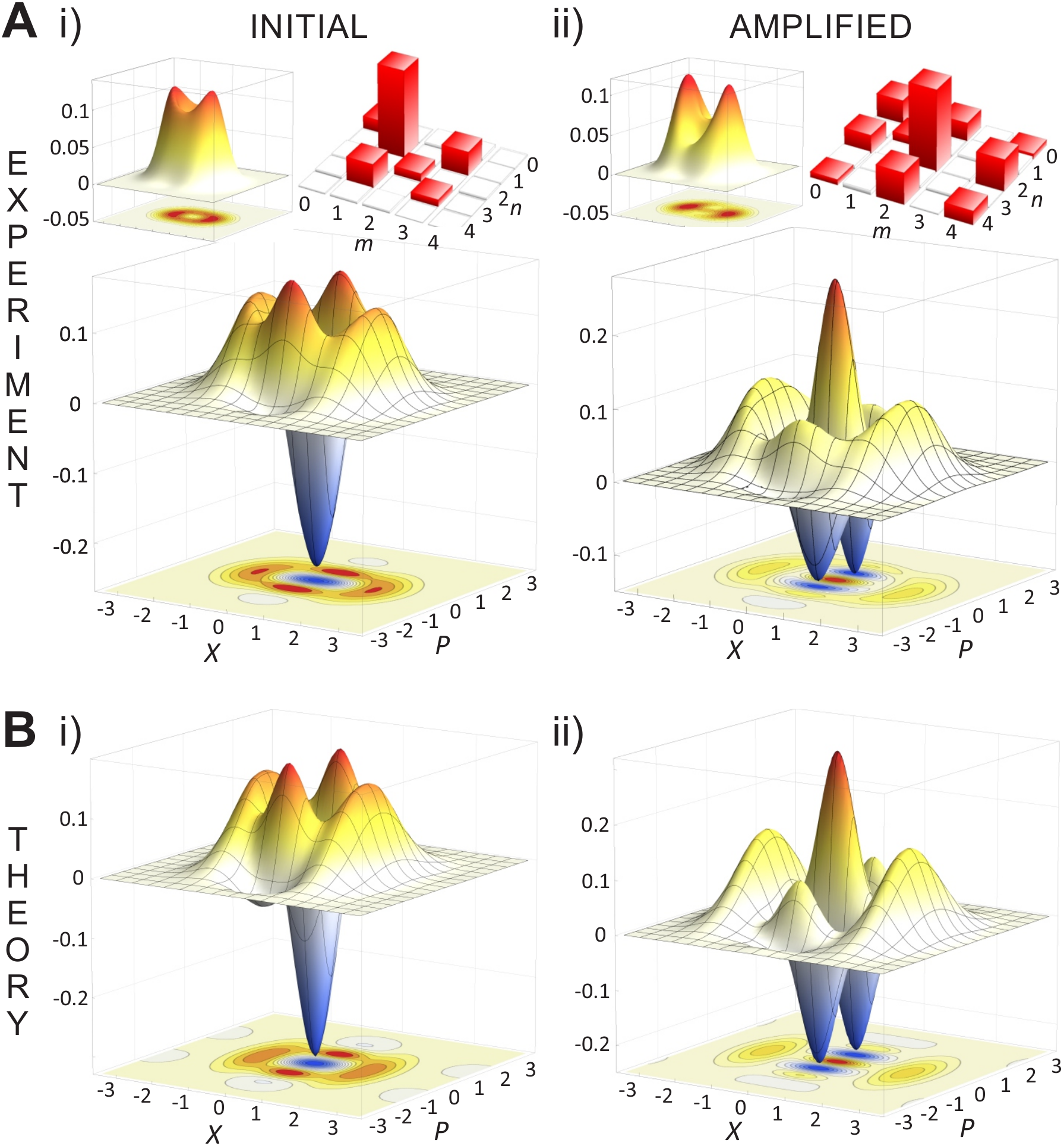}
	\caption{Wigner functions of the initial and amplified SC states. {\bf A}: Experimental reconstruction via homodyne tomography corrected for the total quantum efficiency of 50\%. Left insets: Wigner functions reconstructed without efficiency corrections. Right insets: Absolute values of density matrices $\rho_{mn}$ in the Fock basis reconstructed with efficiency correction. {\bf B}: Best fit with the ideal, squeezed SC state.
		Left (i) [right (ii)]: initial [amplified] SC state. The best fit state is $\ket{{\rm SC}_-\left[1.25\right]}$ [$\ket{{\rm SC}_+\left[2.15\right]}$] squeezed by 1.73 dB [3.47 dB]. The fidelity between the theoretical and experimental [corrected] states is 93\% [86\%]\footnote{If the efficiency correction is performed to compensate only the 62\% losses of the state detection (see Methods), the best fit state is $\ket{{\rm SC}_-\left[1.15\right]}$ [$\ket{{\rm SC}_+\left[1.85\right]}$] squeezed by 2.6 dB [4.8 dB]. The fidelity between the theoretical and experimental (corrected) states is then 78\% [75\%].}.\label{f2}} 
	\end{figure}

Simultaneous preparation of the two initial SC states is heralded by coincident clicks of SPCMs 1 and 2. The states are then  mixed on a symmetric beamsplitter. The optical phase difference between them is actively stabilized (see Methods) to make sure that the output state of the beamsplitter is described by \eeqref{eq2}. Subsequently, one of the beam splitter output modes (mode 2) is subjected to homodyne measurement of the position quadrature. Conditioned on a near-zero measurement result ($|X_2|\le 0.3$) in mode 2, the state of mode 1 is subjected to quantum tomography by means of another homodyne detector. Out of total 40,000 SPCM coincidence events collected, 8,000 satisfied this condition, corresponding to the protocol's success probability of $p=0.2$.

The tomographic reconstruction result is shown in Fig.~\ref{f2}[A(ii)]. The amplified state, corrected for the total efficiency of the initial states, has a fidelity of 86\% with $\ket{{\rm SC}_+\left[2.15\right]}$ squeezed by 3.47 dB displayed in Fig.~\ref{f2}[B(ii)]. Remarkably, the ``breeding" operation does not significantly affect the SC's fidelity. However, the non-ideality of the homodyne detection in mode 2, as well as the imperfection of the initial SC, result in additional squeezing of that state.

The Wigner functions of SC states have a characteristic shape, consisting of two positive Gaussian peaks associated with the individual coherent state constituents and a highly nonclassical ``interference fringe" pattern between them. Our observations are consistent with this description. In the initial SC states, the Gaussian peaks are quite close to each other, so the Wigner function of the initial state resembles that of the squeezed single photon \cite{Ourjoumtsev2006}. For the amplified SC, the peaks are separated further, so one can more clearly distinguish them from the interference pattern in between, with the latter becoming more prominent. This effect on the Wigner function is also quite evident without the efficiency correction (Fig.~2, left insets).



The  protocol demonstrated here constitutes an instrument to convert a pair of  SC states to a single larger-amplitude, positive SC state. The probability of success $p$ is directly related to the width of the  quadrature selection band in output mode 2 of the beam splitter, and asymptotically increases to $1/2$ for high amplitudes. As shown, the fidelity of the amplified SC does not significantly decrease with respect to that of the initial one, permitting the application of the protocol in an iterative fashion. 

A single realization of our protocol produces optical SCs with amplitudes that are comparable to the highest ever achieved, including other physical systems, such as microwave \cite{haroche08} and circuit \cite{vlastakis2013} cavity quantum electrodynamics settings. Iterating our protocol for $n$ stages will further increase the SC amplitude by factor $\alpha'/\alpha=2^{n/2}$. Since each implementation of the protocol would require two input SCs, a total of  $1+2+\ldots+2^{n-1}=2^n-1$ implementations are needed for $n$ stages, with the corresponding success probability of $p^{2^n-1}$. For example, amplifying the SC state from $\alpha=1.4$ to $\alpha'=4$ would require 3 stages and, assuming $p=0.2$, $p^{-2^n}=p^{-(\alpha'/\alpha)^2}\sim4\times 10^5$ copies of the initial states per one copy of the output. However, the use of optical quantum memory in a setting similar to the quantum repeater \cite{QMem} will change the scaling of the overall success probability with respect to the target amplitude from exponential to polynomial. 

\textbf{Methods.}
Degenerate parametric down-conversion takes place in periodically-poled potassium titanyl-phosphate crystals (PPKTP, Raicol), under type-I phase-matching conditions. Each of the two crystals is pumped with $\sim 25$mW frequency-doubled radiation of the master laser, Ti:Sapphire Coherent Mira 900D, with a wavelength of 780 nm, repetition rate of 76 MHz and pulse width of $\sim$1.5 ps. In each nonlinear crystal, a 1.7 dB single-mode squeezed vacuum state is generated \cite{TwoCrSq}.

For the preparation of the initial CSs, 10\% of the energy from these squeezed vacuum states is ``tapped" by beamsplitters and directed to single-photon counting modules (SPCMs, Excellitas) via fiber interfaces. The preparation rate of each SC state is $\sim 10$kHz, which results in a $\sim2$ Hz rate of coincidence events. 

The setup requires two phase-lock loops (PLLs): first, to keep the input SCs in phase with each other and second, to ensure that the homodyne detector in mode 2 measures the $X$ quadrature. To generate the feedback signal, both PLLs use the data from the homodyne detectors without conditioning on the SPCM events (to which we refer as ``non-triggered"). The first PLL is set to minimize the Einstein-Podolsky-Rosen-type quadrature correlations, which show up in the detectors' measurements when the phases are misaligned \cite{TwoCrSq}. The feedback signal is applied to a piezoelectric transducer in one of the initial state's paths, (PZT$_1$ in Fig.~\ref{f1}). 

If the first PLL functions properly, the non-triggered output of the beam splitter constitutes two unentangled single-mode momentum-squeezed states. The second PLL can therefore be set to keep the variance of the mode 2 quadrature measurements at maximum. The feedback signal is applied to the corresponding local oscillator phase wia PZT$_2$ (Fig.~\ref{f1}).

The  reconstruction of both the initial and amplified SC states is performed using the iterative maximum-likelihood algorithm \cite{Lvo2009, Lvo2004}. The local oscillator phase is varied by PZT$_3$ and its time-dependent value is extracted from the variance of the non-triggered quadrature data, which corresponds to the single-mode squeezed state.
 To perform tomography of the initial SCs, the reflectivity of the central beam splitter is set to zero.

The total quantum efficiency of state detection, 50\%, is determined from the analysis of these SC states \cite{Berry10}. The efficiency reduction occurs during both the preparation and detection of the cat states. The preparation losses  include those on the tapping beam splitter (90\% transmissivity) and the imperfect matching between the modes detected by the SPCMs and the
squeezed mode (90\%). The detection imperfections include linear losses on various optical elements (89\% total transmissivity), imperfect mode matching between the signal and local oscillator (81\%) and the efficiency of the homodyne detector \cite{Kumar2012} (86\%).

\noindent
\textit{Acknowledgments.}
We thank Y. Kurochkin and A. Turlapov for discussions.

\end{document}